\documentclass[preprint,notoc]{JHEP3}
\usepackage{cite}
\usepackage{epsfig}
\usepackage{multirow}

\def\etmiss{E_T^{\rm miss}}
\def\eslt{E_T^{\rm miss}}

\def\to{\rightarrow}

\def\bi{\begin{itemize}}
 \def\ei{\end{itemize}}
\def\te{\tilde e}
\def\shat{\hat{s}}
\def\c1p{C1^\prime}
\def\ta{\tilde a}
\def\tG{\tilde G}

\def\tu{\tilde u}

\def\ta{\tilde a}

\def\tb{\tilde b}

\def\tst{\tilde t}

\def\tg{\tilde g}

\def\tell{\tilde\ell}
\def\tq{\tilde q}

\def\tw{\widetilde W}
\def\tz{\widetilde Z}
\newcommand\sjp[3]{{\it Sov.\ J.\ Nucl.\ }{\bf #1} (#2) #3}
\def\alt{\lesssim}
\def\agt{\gtrsim}
\def\be{\begin{equation}}  
\def\ee{\end{equation}}  
\def\bea{\begin{eqnarray}}  
\def\eea{\end{eqnarray}}  

\def\sps1ap{SPS1a$^\prime$}
\title{Testing Yukawa-unified SUSY during year 1 of LHC:\\
the role of multiple $b$-jets, dileptons and missing $E_T$}
\author{Howard Baer$^{a}$, Sabine Kraml$^{b}$, Andre Lessa$^{a}$
and Sezen Sekmen$^{c}$ \\
$^a$Dept.\ of Physics and Astronomy, University of Oklahoma, Norman, OK 73019, USA\\
$^b$Laboratoire de Physique Subatomique et de Cosmologie, UJF Grenoble 1, 
CNRS/IN2P3, INPG, 53 Avenue des Martyrs, F-38026 Grenoble, France\\
$^c$Dept.\ of Physics, Florida State University, Tallahassee, FL 32306\\
E-mail: \email{baer@nhn.ou.edu}, 
\email{sabine.kraml@lpsc.in2p3.fr}, 
\email{lessa@nhn.ou.edu},
\email{sezen.sekmen@cern.ch}}

\preprint{\vbox{LPSC09171}}

\abstract{
We examine the prospects for testing SO(10) Yukawa-unified 
supersymmetric  models during the first year of LHC running at
$\sqrt{s}= 7$ TeV, assuming integrated luminosity values of
$\sim$ 0.1--1~\rm fb$^{-1}$. We consider two cases: the
Higgs splitting (HS) and the $D$-term splitting (DR3) models.
Each generically predicts light gluinos and heavy squarks, with an inverted scalar  
mass hierarchy. We hence expect large rates for gluino pair production 
followed by decays to final states with large $b$-jet multiplicity. 
For $0.2$~fb$^{-1}$ of integrated luminosity, we find a $5\sigma$ 
discovery reach of $m_{\tg}\sim 400$ GeV even if missing transverse energy, 
$\eslt$, is not a viable cut variable, by examining the multi-$b$-jet final state.
A corroborating signal should stand out in the opposite-sign (OS)
dimuon channel in the case of the HS model; the DR3 model will require
higher integrated luminosity to yield a signal in the OS dimuon channel.
This region may also be probed by the Tevatron with 5--10~fb$^{-1}$ 
of data, if a corresponding search in the multi-$b+\eslt$ 
channel is performed. 
With higher integrated luminosities of $\sim 1$~fb$^{-1}$, using $\eslt$ plus
a large multiplicity of $b$-jets, LHC should be able to 
discover Yukawa-unified SUSY with $m_{\tg}\alt 630$ GeV.
Thus, the year 1 LHC reach for Yukawa-unified SUSY should be enough
to either claim a discovery of the gluino, or to very nearly rule out this 
class of models, since higher values of $m_{\tg}$ lead to rather poor 
Yukawa unification.
}  

\begin{document}

\section{Introduction}
\label{sec:intro}

Grand unified theories (GUTs) find a welcome inclusion of supersymmetry
(SUSY) into their structure in that SUSY tames the gauge hierarchy problem
via the well-known cancellation of quadratic divergences\cite{review}.
In particular, the GUT group $SO(10)$ is highly motivated
in that it allows for-- in addition to gauge unification-- the unification
of all the matter superfields of each generation into the 16-dimensional
spinor representation\cite{so10}. The matter unification only works if the 15 matter 
superfields of the Minimal Supersymmetric Standard Model (MSSM) 
are augmented by a SM gauge singlet superfield $\hat{N}_i^c$ which 
contains a right-hand neutrino (RHN) field. 
The presence of RHN fields is essential to describe data from the past
decade on neutrino mass and flavor oscillations; in particular
a Majorana mass term near the GUT scale, needed to implement
see-saw neutrino masses\cite{seesaw}, should be generated by the
breakdown of $SO(10)$ gauge symmetry. 
In addition to gauge and matter unification, in the simplest
$SO(10)$ SUSY GUT models-- wherein both MSSM Higgs doublets
reside in a 10 of $SO(10)$-- one expects {\it Yukawa coupling unification}
in the third generation: $f_t=f_b=f_\tau \ (=f_{\nu_\tau})$ at $M_{\rm GUT}$.

Recently, a variety of studies have examined the MSSM(+RHN) 
to check whether the measured values of gauge couplings and third generation
fermion masses do indeed allow for $t-b-\tau$ Yukawa coupling 
unification\cite{early,also,bdft,bf,bdr,abbbft,bdGaugino,bkss,agrs,gks,grs,dr3}.
Essential to the calculation is the inclusion of 2-loop renormalization
group equations\cite{mv} (RGEs) and inclusion of 
weak scale threshold corrections\cite{hrs}
which occur due to the $\rm MSSM\to SM$ transition in effective field theories.
These threshold corrections imply that Yukawa coupling unification depends
on the entire spectrum of SUSY particles, since the SUSY particles enter the
various $t$, $b$ and $\tau$ self-energy diagrams\cite{hrs}.

Assuming universal boundary conditions at the GUT scale, 
the parameter space of $SO(10)$-motivated SUSY consists of
\be
m_{1/2},\ m_{16},\ m_{10},\ M_D^2,\ A_0,\ \tan\beta ,\ sign(\mu ) ,
\label{eq:pspace}
\ee
where $m_{1/2}$ is the common gaugino mass at $M_{\rm GUT}$, 
$m_{16}$ is the common GUT mass of all matter scalars,
$m_{10}$ is that of the Higgs soft terms, and $M_D^2$
parametrizes potential splittings in the GUT scale Higgs (and possibly matter scalar) 
soft terms. Such splittings are expected to arise from the breaking of the $SO(10)$.
It has been found that $t-b-\tau$ Yukawa coupling unification can occur in
the MSSM within this setup, but only for very restricted forms of the soft SUSY 
breaking (SSB) parameters at $M_{\rm GUT}$. 
These include, for the case of $\mu >0$:
\bi
\item $A_0^2=2m_{10}^2=4m_{16}^2$,
\item $m_{16}\sim 5-15$ TeV,
\item $m_{1/2}\ll m_{16}$,
\item $\tan\beta\sim 50 $.
\ei
These boundary conditions were found in Ref.~\cite{bfpz} to give rise
to an inverted scalar mass hierarchy (IMH), wherein first/second 
generation scalars end up with masses $\sim 10$ TeV, while third 
generation scalars, Higgs scalars $A,\ H$ and $H^\pm$ and $\mu$
are of order $\sim 1-2$ TeV.

A problem with the IMH scheme is that it is inconsistent with
radiative electroweak symmetry breaking (REWSB), unless
the Higgs soft terms are split at $M_{\rm GUT}$\cite{murayama}: 
$m_{H_u}^2<m_{H_d}^2$, thus giving $m_{H_u}^2$ a head start 
over $m_{H_d}^2$ in its running towards the weak scale.\footnote{This 
can be different in non-universal models, see \cite{bdGaugino,gks,grs}.} 
Such splitting naturally occurs due to $D$-term (DT) contributions to {\it all} 
scalar masses arising from the breakdown of $SO(10)$. 
However, applying the splitting
to {\it only} the Higgs sector (``just-so'' Higgs splitting, HS) 
\be
m_{H_{u,d}}^2= m_{10}^2\mp 2M_D^2 \ \ \ \ \ ({\rm HS\ model})
\ee
results in better accuracy of Yukawa unification as compared to 
full DT splitting.
Recently, it has been shown that DT splitting, combined with the 
running effect of the neutrino Yukawa coupling $f_{\nu_\tau}$ and
a small mass splitting between first/second versus third generation scalars
(the DR3 model) can allow for Yukawa coupling unification to a few percent\cite{dr3}.

Both the HS and DR3 schemes lead to sparticle mass spectra characterized by
\bi
\item $m_{\tq ,\tell}(1,2)\sim 10$ TeV,
\item $m_{\tq ,\tell}(3)\ {\rm and}\ \mu\sim 1-3$ TeV,
\item $m_{\tg}\sim 300-500$ GeV,
\item $m_{\tw_1 ,\tz_2}\sim 100-180$ GeV,
\item $m_{\tz_1}\sim 50-90$ GeV.
\ei
Figure~\ref{fig:mgR} shows the location of a large number of 
Yukawa-unified models in the $R\ vs.\ m_{\tg}$ plane, for the HS model
(red dots) and the DR3 model (blue dots) obtained through a Markov Chain Monte Carlo (MCMC) scan of the
parameter space (for details, see\cite{dr3}). 
Here, the degree of Yukawa unification is quantified as
\begin{equation}
R=\frac{max(f_t,\ f_b,\ f_{\tau})}{min(f_t,\ f_b,\ f_{\tau})} 
\label{eq:R}
\end{equation}
where $f_t,\ f_b$ and $f_{\tau}$ are the top, bottom and tau Yukawa couplings, respectively,
evaluated at $Q=M_{\rm GUT}$.
In the DR3 case, if we
require $R<1.05$, then $m_{\tg}\alt 450$ GeV. In the HS model,
while $m_{\tg}\sim 300-500$ GeV is favored for low $R<1.05$ 
solutions, it is possible (but not likely) to have occassional models with
$m_{\tg}$ as large as $\sim 700$ GeV.

\FIGURE[t]{
\includegraphics[width=14cm,clip]{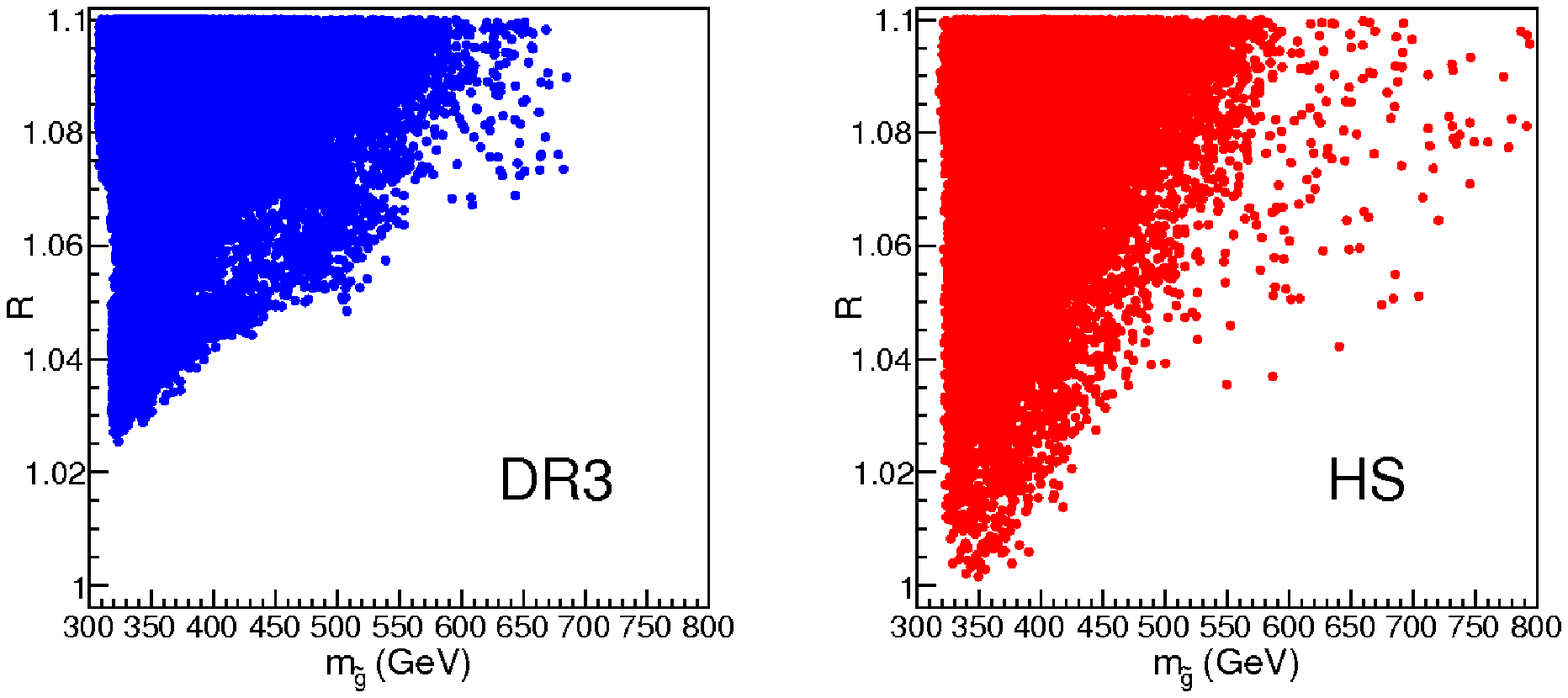}
\caption{Scatter plot of Yukawa unified models in the $R\ vs.\ m_{\tg}$
plane, for solutions in the DR3 model (blue) and the HS model (red).
  }
\label{fig:mgR}}

In models with the above listed superpartner spectrum and a bino-like
$\tz_1$ state, the neutralino relic density is computed to be
$\sim 10^2-10^4$ times the measured abundance\cite{abbbft,bkss},
and the models are seemingly excluded. 
However, if one invokes
the Peccei-Quinn solution to the strong $CP$ 
problem\cite{pq,ww,ksvz,dfsz,axreview}, 
then an axion/axino supermultiplet is expected in the theory\cite{nr}. With an 
axino of mass $m_{\ta}\sim 1$ MeV the neutralinos will decay
via $\tz_1\to \ta\gamma$, which greatly reduces the dark matter density 
by a large factor: $m_{\ta}/m_{\tz_1}$. Cold dark matter (CDM)
solutions can be found consisting of mainly cold axions and thermally
produced axinos, with a small component of warm axinos arising from
$\tz_1\to\gamma\ta$ decay, which occurs on time scales of order $1$ sec.
Since $m_{16}\sim 10$ TeV, and we expect $m_{16}\sim m_{\tG}$, the
axion/axino CDM scenario allows for a solution to the gravitino BBN
problem, and can generate re-heat temperatures $T_R\sim 10^6-10^9$
GeV, which can allow for baryogenesis mechanisms such as
non-thermal\cite{ntlepto} or Affleck-Dine\cite{adlepto} 
leptogenesis to occur\cite{bhkss}.

Since the value of $m_{\tg}$ is so low in Yukawa-unified SUSY models,
we expect the whole scenario to soon be tested at the CERN LHC.\footnote{Indeed 
experiments at the Fermilab Tevatron collider may also probe up to 
$m_{\tg}\sim 400-430$~GeV\cite{tev}.}
LHC has already turned on in Fall, 2009. As time progresses, the 
centre-of-mass energy $\sqrt{s}$ will be increased into the 
$\sim 7$ TeV regime.
An integrated luminosity of $0.1-1~\rm fb^{-1}$ is expected to be collected.\footnote{The 
quoted physics data to be collected at 7 TeV as of November 2009 is 0.2 fb$^{-1}$.}
Earlier work on Yukawa-unified SUSY at LHC with $\sqrt{s}=14$ TeV showed
the model to be easily testable at LHC\cite{so10lhc}. 
The LHC SUSY events should be
characterized by gluino pair production followed by three-body
decays to states including a high multiplicity of $b$-jets. In addition, 
opposite-sign dileptons with mass between $40-80$ GeV ({\it i.e.} between
the $\gamma$ and $Z$ peaks) may be evident.

In the intervening past year, while LHC recovered from an unfortunate
incident involving faulty circuits and quenched magnets, the experiments
have been measuring millions of cosmic muon events. This cosmic data
has allowed them to fine-tune their detector response to muons, and 
to make great strides in alignment of detector elements. We expect
thus that isolated muons, jets and $b$-jets should be readily measurable
very early on during LHC running, while reliable electron identification ($e$ ID) 
and even more so reliable $\eslt$ measurement, may require additional time
to establish.

\clearpage
In this paper, we expand upon the analysis presented in Ref.~\cite{so10lhc}, 
and address several new issues:
\begin{itemize}
\item We focus on the LHC potential to discover or rule out 
Yukawa-unified SUSY during year 1 of running.\footnote{Some additional analyses 
of early physics prospects at LHC are contained in Refs.~\cite{mlm,earlylhc,earlylhc2}.} 
To this end, we calculate
signal and background production rates for the LHC turn-on energy of
$\sqrt{s}=7$ TeV rather than the maximal collider energy of
$\sqrt{s}=14$ TeV used earlier. 
Part of the effect of LHC turn-on at lower than expected energies can be gleaned from 
Fig.~\ref{fig:xsvsrs}, where we plot $\sigma (pp\to\tg\tg X)$ vs. collider
energy $\sqrt{s}$, for $m_{\tg}=300,\ 400$ and 500 GeV, while taking 
$m_{\tq}=10$ TeV. We show both LO and NLO QCD results as derived from Prospino\cite{prospino}.
For $m_{\tg}=400$ GeV, LHC operating at $\sqrt{s}=7$ TeV yields a
cross section of $\sigma\sim 10^4$ fb. 
It is expected that, after about $0.1~{\rm fb}^{-1}$ of integrated luminosity, the LHC will 
move up in energy to the $\sqrt{s}\sim 10$ TeV regime, where $\sigma\sim 3\times 10^4$ fb. 
Ultimately, the LHC should move up to its design energy of $\sqrt{s}=14$ TeV,
where the cross section increases to $\sim 10^5$ fb.
Various background rates will also change accordingly. In this paper, we take a conservative approach,
and evaluate all signal and background cross sections at $\sqrt{s}=7$ TeV. Increasing the beam energy
beyond 7 TeV should only increase the SUSY reach projections which we calculate here.
\FIGURE[t]{
\includegraphics[width=11cm,clip]{gluinoXsec.eps}
\caption{ Total cross-section for gluino pair production
with $m_{\tq}=10$ TeV versus LHC collider energy $\sqrt{s}$, 
for $m_{\tg}=300$, 400 and 500 GeV.
}
\label{fig:xsvsrs}}
\item We include as well many more background subprocesses
than before, including the effect of many $2\to 3$ and $2\to 4$ body 
subprocesses. 
\item We particularly hone in on what LHC can accomplish
with very low integrated luminosity. After turn-on, some time will be required
to examine detector response to well-known SM processes like $W$, $Z$
and $t\bar{t}$ production. To be able to use the classic SUSY signature of
$jets+\eslt$ production, the measurement of $\eslt$-- which depends
on a knowledge of the entire detector response-- will be required.
However, in Ref.~\cite{earlylhc,earlylhc2}, it is pointed out that LHC experiments
can examine multi-jet $+$ isolated multi-muon events in lieu of
$jets+\eslt$ events as a gain for signal over background.
We find that for very low integrated luminosity, using either
large isolated muon multiplicity, or large $b$-jet multiplicity, allows
$m_{\tg}$ values of up to 400 GeV to be probed with just 0.2 fb$^{-1}$
of integrated luminosity. (Note that we expect a similar reach for the Tevatron 
in the $\ge 2-3~b$-jets $+\,\eslt$ channel with $5-10~\rm fb^{-1}$\cite{tev}.)
\item We also discuss the case when $\eslt$ measurements and $e$ ID are established. 
Here we find that LHC can explore $m_{\tg}$ values as high as $\sim 630$ GeV 
with 1~fb$^{-1}$ of integrated luminosity. Thus, during year 1 the LHC 
may well be able to either discover or very nearly rule out Yukawa-unified SUSY.
\end{itemize}
 
The paper is organized as follows. 
We first establish in Sec. \ref{sec:pheno} two
Yukawa-unified model lines: one in the HS model and one in the DR3 
model. We also examine general features in sparticle production and decay
for these model lines. In Sec. \ref{sec:sim},
we present some technical details of our signal and background calculations.
In Sec. \ref{sec:early}, we present expectations for {\it early} SUSY
searches in the multi $b$-jets channel\footnote{Earlier work emphasizing 
the utility of the presence of $b$-jets in SUSY events was 
provided in Refs.~\cite{xt,so10lhc}.} {\it without} using $\eslt$ cuts. 
We also examine rates for early multi-muon production plus jets 
without using  $\eslt$.
We find that the HS and DR3 models may be distinguishable
by measuring the ratio of OS dilepton events to multi $b$-jet events,
since both models produce multi-$b$-jets at a similar rate. However,  
while OS dimuons from $\tz_2$ decay are abundant in the HS model, they
are relatively scarce in the DR3 model.
In Sec. \ref{sec:reach}, we move beyond the 0.1~fb$^{-1}$ level, 
and calculate the LHC reach for the two model lines using as well
$\eslt$ and $e$ ID for 1~fb$^{-1}$ of integrated luminosity.
In this case, the $5\sigma$ LHC reach should extend to $m_{\tg}\sim 630$~GeV, 
enough to cover the bulk of parameter space of these simple Yukawa-unified models.
In Sec. \ref{sec:conclude}, we present our conclusions.

\section{HS and DR3 model lines}
\label{sec:pheno}

\subsection{Model lines}

Using the parameter space in Eq.~\ref{eq:pspace}, Ref.~\cite{bhkss} found
a large number of SUSY spectral solutions with good Yukawa coupling unification 
in the HS model. 
We adopt Point B with $R=1.02$ of this paper as a Yukawa-unified benchmark 
point, and label it as HSb.
The HSb input parameters and mass spectra are listed in Table \ref{tab:BM}.
%
\begin{table}\centering
\begin{tabular}{lcc}
\hline
parameter & HSb & DR3b \\
\hline
$m_{16}(1,2)$& 10000 & 11805.6 \\
$m_{16}(3)$  & 10000 & 10840.1 \\
$m_{10}$   & 12053.5 & 13903.3 \\
$M_D$      & 3287.1 & 1850.6 \\
$m_{1/2}$  & 43.9442 & 27.414 \\
$A_0$      & $-19947.3$ & $-22786.2$ \\
$\tan\beta$& 50.398 & 50.002 \\
$R$        & 1.025  & 1.027 \\ 
$\mu$      & 3132.6 & 2183.4  \\
\hline
$m_{\tg}$   & 351.2 & 321.4   \\
$m_{\tu_L}$ & 9972.1 & 11914.2   \\
$m_{\tst_1}$& 2756.5 & 2421.6   \\
$m_{\tb_1}$ & 3377.1 & 1359.5  \\
$m_{\te_R}$ & 10094.7 & 11968.5  \\
$m_{\tw_1}$ & 116.4 & 114.5  \\
$m_{\tz_2}$ & 113.8 & 114.2  \\ 
$m_{\tz_1}$ & 49.2 &  46.5  \\ 
$m_A$       & 1825.9 &  668.3  \\
$m_h$       & 127.8 &  128.6  \\ 
\hline
\end{tabular}
\caption{Masses in~GeV units and parameters
for Yukawa-unified benchmark points HSb~\cite{bhkss} and DR3b~\cite{dr3}.
For the DR3 model, we use $M_{N_3}=10^{13}$ GeV.
}
\label{tab:BM}
\end{table}

To construct a HS model line, we keep most of the above parameters
fixed, but allow $m_{1/2}$ to vary. This keeps the Yukawa-unification
generally low, but allows us to vary 
$m_{\tg}\sim 3.5m_{\tz_2}\sim 7 m_{\tz_1}$ continuously. 
We plot the value of $R$ versus $m_{\tg}$ in Fig.~\ref{fig:R}.
We see that at low $m_{\tg}$ ($\sim 325$ GeV), $R< 1.03$, while
as $m_{\tg}$ increases, Yukawa unification gets worse until
$m_{\tg}\sim 700$ GeV, where we find $R\sim 1.13$.

\FIGURE[t]{
\includegraphics[width=11cm,clip]{Rlamb.eps}
\caption{Degree of Yukawa unification $R$ (see Eq.\ref{eq:R})
for the models HS and DR3 as a function of the gluino mass.
The model parameters are the same as in Table \ref{tab:BM},
but with $m_{1/2}$ varying from 30 to 180 GeV.
  }
\label{fig:R}}

We also adopt from Ref.~\cite{dr3} a DR3 model line, labelled as DR3b,
with parameters in Table \ref{tab:BM}.
where $m_{16}(1,2,3)$, is the scalar mass for the 1st, 2nd and 3th generations and $M_{N_3}$,
$f_{\nu_\tau}$, $A_{\nu_{\tau}}$ and $m_{\tilde{v}_{R3}}$ are the right-handed neutrino mass, Yukawa coupling,
$A$-term and the scalar mass for the sneutrino.
We construct a DR3 model line by again keeping most parameters fixed, but
letting $m_{1/2}$ to vary. In the DR3 model line, we find $R\sim 1.03$ 
for $m_{\tg}\sim 325$ GeV, while $R$ increases to $\sim 1.15$ for
$m_{\tg}\sim 700$ GeV.\footnote{Although Fig.~\ref{fig:mgR} shows that for some special choice of the
parameters lower values of $R$ can be obtained for $m_{\tg}\sim 700$, the model
lines chosen here represent the general behavior of the bulk of parameter space.}

Due to the heavy scalar masses in the HS and DR3 model lines, 
the scalars essentially decouple at LHC energies.
What results is a low energy effective theory where only 
$\tilde{g}$, $\tz_{1,2}$ and $\tw_1^{\pm}$ are 
the new physics matter states. In the next two sections we discuss the
production cross-sections and decay rates for these states.

\subsection{HS and DR3 production cross sections}

We plot in Fig.~\ref{fig:xsecs} the leading order $\tg\tg$, $\tw_1\tz_2$ and 
$\tw_1^+\tw_1^-$ production cross sections as a function of
$m_{\tg}$ for collider energy $\sqrt{s}=7$ TeV. 
From the figure, we see that gluino pair production is  dominant up to 
$m_{\tilde{g}} \sim$ 520 GeV (at NLO, it dominates up to 
$m_{\tilde{g}} \sim$ 560 GeV), with cross sections 
typically greater than $10^3$ fb, and in excess of $10^4$ fb in the
lower gluino mass range. Thus, even with integrated luminosities
as low as $0.1$ fb$^{-1}$, we expect hundreds of gluino pair events
in the upcoming LHC year 1 physics data sample for the HS and DR3 models.

Gluino pair production cross sections at the Fermilab Tevatron collider
show a large increase in rate as $m_{\tq}$ increases\cite{tev}. This is due to
suppression of negative interference terms in the $q\bar{q}\to \tg\tg$ subprocess
cross section. At the LHC, gluino pair production for $m_{\tg}\sim 300-500$ GeV 
is dominated instead by the $gg\to\tg\tg$ subprocess, which is independent of
$m_{\tq}$. Thus, $\tg \tg$ cross sections show only a slight ($\sim$10--20\%) 
increase with increasing $m_{\tq}$ at the LHC.

\FIGURE[t]{
\includegraphics[width=11cm,clip]{xsecs.eps}
\caption{ Leading order total cross-sections for sparticle production
in the HS model (solid) and DR3 model (dashed) as a function
of the gluino mass for $pp$ collisions at $\sqrt{s}=7$ TeV.
The model parameters are the same as in Table \ref{tab:BM}
but with $m_{1/2}$ varying from 30 to 180 GeV.
}
\label{fig:xsecs}}

\subsection{Sparticle branching fractions in the HS and DR3 model lines}

Since $m_{\tg}\ll m_{\tq}$ in the HS or DR3 model lines, we will get 
dominant gluino decays into three-body modes.
The gluino branching ratios will be largely model dependent, 
but since $\tilde{t}_i$ and $\tilde{b}_i$ are always the lightest squarks, 
and $\tan\beta$ is large\cite{bcdpt}, 
the decays will mostly be restricted to the following channels:

\begin{itemize}
\item $\tilde{g}\to\tz_i+b\bar{b}$, $i=1,2$
\item $\tilde{g}\to\tz_1+t\bar{t}$
\item $\tilde{g}\to\tw_1^-\bar{b}t\ or\ \tw_1^+b\bar{t}$ .
\end{itemize}

The general feature $m_{\tg}\ll m_{\tq}$ is common to both the HS and DR3 models, 
since it relies mostly on the fact that $m_{1/2} \ll m_{16}$. 
However, the inclusion of the $D$-term splitting for all matter scalars in the
DR3 model pushes $m_{\tilde{b}_R}$ to lower values, when compared to the HS model,
where $m_{\tilde{b}_L} \sim m_{\tilde{b}_R}$.\footnote{The stop masses and mixing are basically
the same in both models, 
since the $D$-term splitting is equal for both $\tilde{t}_R$ and $\tilde{t}_L$.}
As a result we have:
\begin{itemize}
\item DR3: $\tilde{b}_1 \sim \tilde{b}_R$ and $m_{\tilde{b}_1} < m_{\tilde{b}_2}$
\item HS: $\tilde{b}_1 \sim \tilde{b}_L$ and $m_{\tilde{b}_1} \sim m_{\tilde{b}_2}$
\end{itemize}
Now, since $\tz_2$ is wino-like in both models, it just couples to left-squarks, what 
suppresses the $\tilde{g} \to \tz_2 + b\bar{b}$ decay in the DR3 model and favors  
it in the HS case.
This behavior is shown in Fig.~\ref{fig:brs}, where the main gluino branching ratios 
for both models are plotted as a function of the gluino mass. 
From Fig.~\ref{fig:brs} it can also be seen that-- in the HS model-- 
once $m_{\tilde{g}} \gg m_{t}+m_{\tw_1}$
the $\tilde{g} \to \tw_1^- \bar{b}t+c.c.$ channel starts to dominate, 
since $m_{\tilde{t}_1} < m_{\tilde{b}_1}$ (for $m_{\tilde{g}}>$ 500 GeV).
We can also see that $\tilde{g}\to\tz_1 t\bar{t}$ becomes relevant for heavy gluinos
($m_{\tilde{g}} > 600$ GeV) and it is enhanced in the DR3 model, 
where $\tilde{t}_1$ is usually lighter than in the HS model.

\FIGURE[t]{
\includegraphics[width=7.cm,clip]{hs_brs.eps}\hspace{.5cm}
\includegraphics[width=7.cm,clip]{dr3_brs.eps}
\caption{Gluino branching ratios for the HS and DR3 model-lines as a function
of the gluino mass.
The model parameters are the same as in Table \ref{tab:BM},
but with $m_{1/2}$ varying from 30 to 180 GeV.
  }
\label{fig:brs}}

From Fig.~\ref{fig:brs} we see that, for the HSb benchmark case:
\bi
\item $BR(\tilde{g}\to\tz_2+b\bar{b})=$ 63$\%$
\item $BR(\tilde{g}\to\tz_1+b\bar{b})=$ 15$\%$
\item $BR(\tilde{g}\to\tw_1 + bt)=$ 9$\%$ .
\ei
On the other hand, the DR3b point has:
\bi
\item $BR(\tilde{g}\to\tz_2+b\bar{b})=$ 11$\%$
\item $BR(\tilde{g}\to\tz_1+b\bar{b})=$ 86$\%$
\item $BR(\tilde{g}\to\tw_1 + bt)=$ 0.3$\%$ .
\ei

The $\tz_2$ and $\tw_1$ are expected to decay via three body modes:
\begin{itemize}
\item $\tz_2\to\tz_1 f\bar{f}$
\item $\tw_1^{\pm}\to\tz_1 f\bar{f}'$ ,
\end{itemize}
where the decays are dominated by the intermediate virtual $W^*$ and $Z^*$
diagrams. If $m_{\tg}\agt 500$ GeV, then the two-body modes
$\tw_1\to\tz_1 W$ and $\tz_2\to\tz_1 Z$ will turn on.

\FIGURE[t]{
\includegraphics[width=7.cm,clip]{hs_brs2.eps}\hspace{.5cm}\includegraphics[width=7.cm,clip]{dr3_brs2.eps}
\caption{$\tz_2$ branching ratios for the HS and DR3 model-lines as a function
of the gluino mass.
The model parameters are the same as in Table \ref{tab:BM},
but with $m_{1/2}$ varying from 30 to 180 GeV.
  }
\label{fig:bfz2}}

Putting all segments of the cascade decays together, 
we expect the HSb signal to be rich in $b$-jets and 
opposite-sign/same-flavor (OS/SF) isolated dileptons 
coming from $\tz_2 \to \tz_1 \ell\bar{\ell}$, with
a small rate of SS dileptons coming from $\tg\to\tw_1 q\bar{q}'$ followed by
$\tw_1\to\ell\nu_{\ell}\tz_1$ decay. 
For the DR3b point, we expect the signal to be rich in $b$-jets 
with a harder $\etmiss$ spectrum (when compared to
HSb) due to the direct gluino decay to $\tz_1$, 
but with small rates in the multilepton channels.

\section{Event Simulation}
\label{sec:sim}

In order to study the discovery potential of the LHC at $\sqrt{s}=7$ TeV 
we used AlpGen\cite{alpgen} and MadGraph\cite{madgraph} to generate the 
background hard scattering events and Pythia\cite{pythia} for the
subsequent showering and hadronization.  Table~\ref{table:bgs} lists the $2 \to n$ subprocesses included in this study where jets = $u,d,s,c$ and $g$. 
For all the processes involving multiple jets, the MLM matching algorithm\cite{alpgen} was
used to avoid double counting. 
All the above processes were generated at LO, but a relatively low
renormalization and factorization scale ($Q=\sqrt{\shat}/6$)
was used to bring the total cross-sections closer to their NLO values (for more details see Ref.\cite{earlylhc2}).
The signal events were generated using Isajet 7.79\cite{isajet}.

\begin{table}
\centering
\begin{tabular}{|l|c|c|c|}
\hline
                    &                  & Cross   & number of \\
SM process & Generator & section & events \\
\hline
QCD: $2$, $3$ and $4$ jets ($p_T>40$ GeV) & AlpGen & $3.0\times 10^9$ fb  & 13M\\
$t\bar{t}$: $t\bar{t}$ + 0, 1 and 2 jets & AlpGen & $1.6\times 10^5$ fb&  5M\\
$b\bar{b}$: $b\bar{b}$ + 0, 1 and 2 jets & AlpGen & $8.8\times 10^7$  fb&  91M\\
$Z$ + jets: $Z/ \gamma (\to l\bar{l},\nu \bar{\nu})$ + 0, 1, 2 and 3 jets & AlpGen & $8.8\times 10^6$ fb&  13M\\
$W$ + jets: $W^{\pm} (\to l\nu)$ + 0, 1, 2 and 3 jets & AlpGen & $1.8\times 10^7$ fb&  19M\\
$Z$ + $t\bar{t}$: $Z/ \gamma (\to l\bar{l},\nu\bar{\nu})$ + $t\bar{t}$ + 0, 1 and 2 jets & AlpGen & $53$ fb &  0.6M\\
$Z$ + $b\bar{b}$: $Z/ \gamma (\to l\bar{l},\nu\bar{\nu})$ + $b\bar{b}$ + 0, 1 and 2 jets & AlpGen & $2.6\times 10^3$ fb  &  0.3M\\
$W$ + $b\bar{b}$: $W^{\pm} (\to l\nu)$ + $b\bar{b}$ + 0, 1 and 2 jets & AlpGen & $6.4\times 10^3$ fb &  9M\\
$t\bar{t}t\bar{t}$ & MadGraph & $0.6$ fb &  1M\\
$t\bar{t}b\bar{b}$  & MadGraph & $1.0\times 10^2$ fb &  0.2M\\
$b\bar{b}b\bar{b}$ & MadGraph & $1.1\times 10^4$ fb &  0.07M\\
\hline
\end{tabular}
\caption{Background processes included in this study, their cross sections and number of generated events.}
\label{table:bgs}
\end{table}

A toy detector simulation is then employed with calorimeter cell size
$\Delta\eta\times\Delta\phi=0.05\times 0.05$ and $-5<\eta<5$ . The HCAL
(hadronic calorimetry) energy resolution is taken to be
$80\%/\sqrt{E}+3\%$ for $|\eta|<2.6$ and FCAL (forward calorimetry) is
$100\%/\sqrt{E}+5\%$ for $|\eta|>2.6$, where the two terms are combined
in quadrature. The ECAL (electromagnetic calorimetry) energy resolution
is assumed to be $3\%/\sqrt{E}+0.5\%$. We use the Isajet
jet finding algorithm (cone type) to group the hadronic final states
into jets. The jets and isolated lepton definitions are as follow:
\bi
\item Jets are required to have $R\equiv\sqrt{\Delta\eta^2+\Delta\phi^2}\leq0.4$ and $E_T(jet)>25$ GeV.
\item Leptons are considered isolated if they have $p_T(l)>5 $ GeV 
with visible activity within a cone of $\Delta R<0.2$ of $\Sigma E_T^{cells}<5$ GeV.
\ei

Jets are tagged as $b$-jets if they contain a B hadron with $E_T(B)>$ 15 GeV, $\eta(B)<$ 3 and
$\Delta R(B,jet)<$ 0.5. We assume a tagging efficiency of 60$\%$ and light quark and gluon jets can be mis-tagged
as a $b$-jet with a probability 1/150 for $E_{T} \leq$ 100 GeV, 1/50 for $E_{T} \geq$ 250 GeV, 
with a linear interpolation
for 100 GeV $\leq E_{T} \leq$ 250 GeV (see R. Kadala {\it et al.} in Ref.~\cite{xt}).

\section{Early searches without $\eslt$ or electron ID}
\label{sec:early}

During the early stages of data taking at the LHC with integrated luminosity of
order $\sim 0.1$ fb$^{-1}$, it is possible that $\etmiss$ and electron
identification will not be reliable observables (see Refs.~\cite{earlylhc,earlylhc2}). 
Therefore we separate our analysis into two stages:
\bi
\item early searches, where no $\etmiss$ cuts are applied and only 
muons are considered for the leptonic channels, and
\item full analysis, using a minimum $\etmiss$ cut and including both $e$'s and $\mu$'s.
\ei
In both cases we assume that the $b$-jets can be reliably tagged due to their
displaced vertices as reconstructed in the micro-vertex detector or via a non-isolated muon tag.
For initial searches we apply the following set of minimal cuts, labelled C0:\\

\underline{C0 cuts:}
\bi
\item Jet cuts: $n(jets)\geq$ 4 with $E_T(j)\geq$ 50 GeV, $\eta (j) \leq$ 3 and for the hardest jet $E_T(j1)\geq$ 100 GeV,
\item Lepton cuts: $E_T(\ell)\geq$ 10 GeV and $\eta(\ell)\leq$ 2, 
\item $S_T\geq$ 0.2, 
\item $n(b)\ge 1$,
\ei
where $S_T$ is the transverse sphericity and for now $\ell =\mu$ only.

\subsection{Multi $b$-jet signal}

As discussed in Sec.~\ref{sec:pheno} the points HSb and DR3b are expected to be 
rich in $b$-jets and possibly isolated leptons. 
In Fig.~\ref{fig:nbjets} we plot the $b$-jet multiplicity for the signal and backgrounds (BG)
after applying the C0 set of cuts. 
As expected, the BG distribution falls much faster than the signal. 
Signal exceeds BG for $n(b)\geq$ 4, 
where the BG is dominated by multi $b$ production ($b\bar{b}$ and $b\bar{b}b\bar{b})$. 
We also note that the DR3b case has larger rates for 1 $\leq n(b) \leq$ 4 
when compared with the HSb point, due to a lighter gluino. 
However, the HSb benchmark gives a larger signal for $n(b) \geq$ 5, 
since in this case $\tz_2\to\tz_1 b\bar{b}$ also contributes, and $\tz_2$
are produced at large rates from gluino cascade decay.

We note here that a signal rate above an expected SM BG level may not be sufficient to 
claim a discovery, due to both theoretical and experimental uncertainties in 
the multi-$b$ background rate. However, the overall {\it shape} of the $n(b)$
distribution should be to some extent self-normalizing, as one can fix the BG levels
in the $n(b)=0,\ 1,\ 2$ channels, and look for a {\it harder} $n(b)$ distribution in 
the signal case. Thus, some of the uncertainty is removed when one looks for
an excess in {\it ratios} such as $\sigma (n(b)=3)/\sigma (n(b)=1)$.
Table~\ref{table:chan0} shows
the $n(b)\geq$ 3 and 4 cross-sections for both points and the background. 
The signal in the $n(b)\ge 4$ channel is at the 200 fb level and is well
above SM background levels. Such a signal
may be visible with very low integrated luminosity values $\sim 0.05$ fb$^{-1}$! 

\FIGURE[t]{
\includegraphics[width=11cm,clip]{nbjets.eps}
\caption{ $b$-jet distribution after C0
cuts at the LHC, with  $\sqrt{s}=7$ TeV. We show the signal
levels for the HSb (red) and DR3b (green) points along with
various SM backgrounds.
  }
\label{fig:nbjets}}
\begin{table}
\centering
\begin{tabular}{|c|c|c|c|}
\hline
\multicolumn{4}{|c|}{Results after C0-based selection} \\
\hline
 & $\sigma$($n(b)\geq$ 3)  &$\sigma$($n(b)\geq$ 4) &  $\sigma$(OS)   \\
\hline
HSb  & 899 fb & 176 fb  & 99 fb \\ \hline
DR3b  & 1334 fb &  243 fb & 22 fb \\ \hline
BG & 1911 fb & 70 fb & 11 fb\\ \hline
\end{tabular}
\caption[]{Cross-sections for the $n(b)\geq $ 3, 4 and OS channels 
after the C0 cuts for the points HSb, DR3b and the background. }
\label{table:chan0}
\end{table}

As mentioned above, a mere excess in one or more of the multi $b$-jet channels 
may not be sufficient to claim discovery, 
due to large uncertainties in the normalization of the 
high jet multiplicities BGs, such as $b\bar{b}$, $t\bar{t}$ and $Z$ + 2 jets. 
With this in mind,
we present some signal distributions with distinct shapes from the BG ones, 
which could help corroborate a discovery and
provide some information on the sparticle masses.
As discussed in Sec.~\ref{sec:pheno}, events with $n(b) \geq 4$ usually come from 
$\tilde{g} \to \tz_i b\bar{b}$ decays. 
Therefore the invariant mass of the $b\bar{b}$ pair is expected to have edges at 
$m_{\tilde{g}}-m_{\tz_i}$.
This is shown in Figs.~\ref{fig:masshs} and \ref{fig:massdr3}, where 
$max[m_{b_1\bar{b}_1},m_{b_2\bar{b}_2}]$
is plotted\footnote{Here the index $i$ in $b_i$ labels $b$'s coming from the same gluino.} 
at the parton level (dashed black line).
As expected, in the DR3b case the $m_{\tilde{g}}-m_{\tz_1}$
mass edge is much more evident than the $m_{\tilde{g}}-m_{\tz_2}$ edge,
due to the large $\tilde{g} \to \tz_1 + b\bar{b}$ branching fraction,
while the opposite happens for the HSb point. 
At the detector level, the main difficulty in obtaining 
$max[m_{b_1\bar{b}_1},m_{b_2\bar{b}_2}]$ comes from combining the correct $b$-jets into 
pairs coming from the same gluino.
As pointed out in Ref.~\cite{so10lhc}, usually the two hardest
$b$-jets come from different gluinos. 
Furthermore, in most cases the pair coming from the same $\tg$ has smaller separation angles. 
Using these two facts, we select $b_1$ and $b_2$ as the hardest and second hardest jets 
and $\bar{b}_2$ as the remaining jet that minimizes $\Delta\phi (b_2,\bar{b}_2)$. 
Also, to avoid event topologies with no small angular separation, 
we apply the cut $\Delta R(b_2,\bar{b}_2)<1$, where 
$R=\sqrt{\Delta\phi(b_2,\bar{b}_2)^2 + \Delta\eta(b_2,\bar{b}_2)^2}$. 
Using this procedure and adding the C0 set of cuts we obtain the solid curves 
shown in Figs.~\ref{fig:masshs} and \ref{fig:massdr3},
where the BG contribution was added to the signal. We also
show the statistical error bars for 1~fb$^{-1}$ of integrated luminosity.
As can be seen, the invariant-mass distributions have the expected shape, 
although the mass edges seem to require higher integrated luminosity to become statistically relevant.

\FIGURE[t]{
\includegraphics[width=11cm,clip]{massHS.eps}
\caption{Maximum invariant $bb$ mass at parton level for the HSb point without cuts (black/dashed) and
at detector level for the HSb plus background events (red/solid) with $n(b) \geq$ 4, $\Delta R(b_2,\bar{b}_2)<1$
after the C0 cuts (see text).
The BG distribution (gray) and the statistical error bars for 1 fb$^{-1}$ of integrated luminosity are also shown.
  }
\label{fig:masshs}}
\FIGURE[t]{
\includegraphics[width=11cm,clip]{massDR3.eps}
\caption{Maximum invariant $bb$ mass at parton level for the DR3b point without cuts (black/dashed) and
at detector level for the DR3b plus background events (green/solid) with $n(b) \geq$ 4, $\Delta R(b_2,\bar{b}_2)<1$
after the C0 cuts (see text).
The BG distribution (gray) and the statistical error bars for 1 fb$^{-1}$ of integrated luminosity are also shown.
  }
\label{fig:massdr3}}

\subsection{Dimuon channels}

In Fig.~\ref{fig:nmuons} we show the muon multiplicity for the HSb and DR3b signal points and the BG. 
For $n(\mu)= 0,\ 1$, the BG is well above the signal, 
but for $n(\mu)\geq$ 2, signal starts to dominate over the BG.
As expected from the discussion in Sec.~\ref{sec:pheno}, the DR3b
has much smaller rates to multileptons.
Separating the $n(\mu)=$ 2 channel into
opposite sign (OS) and same sign (SS) muons, 
we see that almost all the signal comes from the OS dimuon case.
Due to the small $\tg \to \tw_1^- t\bar{b}+c.c.$ branching fraction, 
the SS signal is almost 2 orders of magnitude below the OS one, 
which makes it irrelevant for luminosities $\lesssim 1$ fb$^{-1}$.
As seen in Table~\ref{table:chan0}, the OS cross-section for the HSb point 
is around 100 fb and
has a discovery potential similar to the multi $b$-jet channel, while
the DR3b benchmark will require more integrated luminosity to be seen in the OS dimuon channel. 

\FIGURE[t]{
\includegraphics[width=11cm,clip]{nmuons.eps}
\caption{ Muon distribution after C0
cuts at the LHC, with  $\sqrt{s}=7$ TeV. We show the signal
levels for the HSb (red) and DR3b (green) points along with
various SM backgrounds. In the $n(\mu)=2$
bin, the left (pluses) and right (crosses) columns show the background (black) and
signal components for OS (with invariant mass cuts) and SS dimuons.  }
\label{fig:nmuons}}

As is well known, the invariant mass of OS muons has a mass edge at
$m_{\tz_2}-m_{\tz_1}$, since most OS muons come from 
$\tz_2 \to \tz_1 \mu^+\mu^-$ decays.
In Fig.~\ref{fig:invmass}, we show the $m(\mu^+\mu^-)$ distribution for the HSb and DR3b points.
As discussed in Sec.~\ref{sec:pheno}, the DR3b point has small leptonic rates 
but may still be visible above background.
In both cases, the mass
edges are visible and give the correct $m_{\tz_2}-m_{\tz_1}$ values. 
We also show the statistical error bars for the
combined signal plus background for 1 fb$^{-1}$ of integrated luminosity. 
From Fig.~\ref{fig:invmass}, we can see that the HSb point gives
a statistically significant edge while the DR3b case may require higher 
integrated luminosities.

\FIGURE[t]{
\includegraphics[width=11cm,clip]{invmass.eps}
\caption{OS dimuon invariant mass for the HSb (red) and DR3b (green) points plus background events
after the C0 cuts (see text).
The BG distribution (gray) and the statistical error bars for 1 fb$^{-1}$ of integrated luminosity are also shown.
  }
\label{fig:invmass}}

Finally, we point out that in both the $n(b) \geq 4$ and OS dimuon channels the 
invariant mass distributions shown in
Figs.~\ref{fig:masshs}, \ref{fig:massdr3} and \ref{fig:invmass}
have distinct features from the BG, which should corroborate a discovery claim.

\subsection{Early SUSY search: reach results}

To estimate the LHC reach in the multi-$b$ and multi-$\mu$ channels, we plot in Fig.~\ref{fig:early_b} 
the signal cross section for the HS and DR3 model lines versus $m_{\tg}$ 
using cuts C0 plus {\it a}).~$n(b)\ge 3$ and {\it b}).~$n(b)\ge 4$, along with the expected BG
rate. We also plot the $5\sigma$ discovery lines for integrated luminosity values 0.1 and 0.2 fb$^{-1}$. The significance in $\sigma$s is derived from the p-value corresponding to the number of S+B events in a Poisson distribution with a mean that equals to the number of background events~\cite{prospercode}.  For both the HS and DR3 model lines, the approximate $5\sigma$ LHC reach extends to
$m_{\tg} \sim 360$ GeV for 0.1 fb$^{-1}$, and $\sim 400$ GeV for 0.2 fb$^{-1}$. 
We remind the reader that this 
is comparable to what Tevatron experiments 
can achieve using $\eslt$ cuts and $> 5$ fb$^{-1}$ of integrated luminosity\cite{tev}.

\FIGURE[t]{
\includegraphics[width=11cm,clip]{mg23b.eps}
\includegraphics[width=11cm,clip]{mg24b.eps}
\caption{Early LHC reach for Yukawa-unified SUSY using cuts $C0$ plus $n_b\ge 3$ and $n_b\ge 4$.
}
\label{fig:early_b}}

The reach using cuts C0 and requiring an OS dimuon pair is shown in Fig.~\ref{fig:early_mu}.
In this case, the reach in the HS model is similar to the multi-$b$ reach:
LHC should explore to $m_{\tg}\sim 360$ GeV with 0.1 fb$^{-1}$, and $m_{\tg}\sim 400$ GeV with
0.2 fb$^{-1}$. In the DR3 model line, however, there is no reach in the OS dimuon channel at these
low values of integrated luminosity. In fact, the rate of multi-$b$-jet events compared to the
rate for OS dimuon events would be one way to distinguish early-on whether one might be
seeing SUSY in the HS or the DR3 model case. 
\FIGURE[t]{
\includegraphics[width=11cm,clip]{mg2OS.eps}
\caption{Early LHC reach for Yukawa-unified SUSY using cuts $C0$ plus 
requiring OS dimuons.
}
\label{fig:early_mu}}

We also point out that despite giving the maximum reach for
both models, the $n(b)\geq 3$ channel has a small signal/BG ratio,
what makes it more dependent on the knowledge of the background.

\section{Analysis including $\eslt$ cut and electron ID}

As the experiments accumulate data, knowledge of the detectors and their response to  
SM background will improve. Also, at some point in time, the LHC center-of-mass energy 
will increase beyond 7 TeV into the 10 TeV range.
To be conservative, we will continue our analysis assuming $\sqrt{s}=7 $ TeV.
Moving to higher values of $\sqrt{s}$ should only increase the possibility of discovering
new, high mass matter states.

As detector response becomes better understood, it will be possible to utilize both 
$\eslt$ and electrons in the analysis. The $\eslt$ variable is well known to be a powerful discriminator between 
SUSY and SM events, and is considered to be the ``classic'' signature for SUSY.  
In Fig.~\ref{fig:etmiss} we show the $\etmiss$ distributions for
the BG as well as the HSb and DR3b points after applying the C0 cuts. 
As expected, the DR3b signal has a harder $\etmiss$ spectrum than HSb, 
due to its large $\tilde{g} \to \tz_1 b\bar{b}$ branching ratio.
\FIGURE[t]{
\includegraphics[width=11cm,clip]{etmiss.eps}
\caption{$\etmiss$ distribution for the BG (gray), HSb (red) and DR3b (green) points after the C0 cuts (see text).
  }
\label{fig:etmiss}}

In order to reduce most of the background and still keep considerably 
large cross-sections for the signal, we henceforth:
\bi
\item include the $\etmiss >$ 100 GeV cut and
\item include $e$'s into the C0 leptonic cuts
\ei
into our analysis. 
The cuts C0 augmented with the $\eslt$ cut and inclusion of $e$'s, but with no
$n(b)$ requirement, will be called C1 cuts:\\

\underline{C1 cuts:}
\bi
\item Jet cuts: $n(jets)\geq$ 4 with $E_T(j)\geq$ 50 GeV, $\eta (j) \leq$ 3 and for the hardest jet $E_T(j1)\geq$ 100 GeV,
\item Lepton cuts: $E_T(\ell)\geq$ 10 GeV and $\eta(\ell)\leq$ 2, 
\item $S_T\geq$ 0.2, 
\item $\etmiss >$ 100 GeV,
\ei
where $\ell =\mu, e$.

\subsection{Multi $b$-jet + $\eslt$ channel}

The main effect of adding an $\etmiss$ cut to our previous analysis is the 
drastic reduction of background in the multi $b$-jet channel. 
However, the signal will also be significantly reduced, and so will be the
statistics in the invariant mass distributions.
\FIGURE[t]{
\includegraphics[width=11cm,clip]{nbjetsB.eps}
\caption{ $b$-jet distribution after C1
cuts at the LHC, with  $\sqrt{s}=7$ TeV. We show the signal
levels for the HSb (red) and DR3b (green) points along with
various SM backgrounds.
  }
\label{fig:nbjetsB}}

The $n(b)$ distribution after C1 cuts is shown in Fig.~\ref{fig:nbjetsB}.
Now the signal's peak at $n(b)=$1, 2 is visible above the BG,  and the hard distribution in
$n(b)$ should be a striking signature for both the DR3b and HSb models, 
since the combined signal plus BG distribution becomes approximately flat for 
0 $\leq n(b) \leq$ 2. 
The $\etmiss$ cut reduces most of the $b\bar{b}$ and $b\bar{b}b\bar{b}$ backgrounds, 
leaving $t\bar{t}$ as the dominant one for $n(b) \geq$ 1. 
This results in a considerable reduction  of the BG in the $n(b)=$ 1, 2 and 3 bins, 
where now the signal/background ratio is larger than one.
The cross-sections for $n(b) \geq$ 3, 4 are shown in Table~\ref{table:chan1}.
Due to the large signal cross-section in these bins, 
the signal could be visible with less than 0.05 fb$^{-1}$ of integrated luminosity
(in the happy case where an $\eslt$ measurement is immediately viable)! 
\begin{table}
\centering
\begin{tabular}{|c|c|c|c|}
\hline
\multicolumn{4}{|c|}{Results after C1-based selection} \\
\hline
 & $\sigma$($n(b)\geq$ 3)  &$\sigma$($n(b)\geq$ 4) &  $\sigma$(OS)   \\
\hline
HSb  & 364 fb & 68 fb  & 81 fb \\ \hline
DR3b  & 782 fb & 139 fb & 23 fb \\ \hline
BG & 16 fb & 2 fb & 9 fb \\
    \hline
\end{tabular}
\caption[]{Cross-sections for the $n(b)\geq $ 3, 4 and OS channels
after the C1 cuts for the points HSb, DR3b and the background. }
\label{table:chan1}
\end{table}

The invariant mass distributions from Figs.~\ref{fig:masshs} and \ref{fig:massdr3}
are now re-plotted in Figs.~\ref{fig:masshsB} and \ref{fig:massdr3B} after the 
$\etmiss$ cut is included. Now, due to the drastic reduction in the background, the
$m_{b\bar{b}}$ distribution is nearly free of BG events. 
However, the decrease in the signal statistics makes it 
impossible to obtain any information on the sparticle masses from the
shape of $max[m_{b_1\bar{b}_1},m_{b_2\bar{b}_2}]$.

A quantity that may still reveal some information on the gluino mass scale 
is the effective mass, $M_{\rm eff}=\sum p_T(jets)+\etmiss$, plotted in 
Fig.~\ref{fig:meffB} for the HSb and DRb points, as well as for two points
on the HS and DR3 model lines with heavier gluinos ($m_{\tilde g}=576$~GeV and 
$m_{\tilde g}=581$~GeV, respectively).

\FIGURE[t]{
\includegraphics[width=11cm,clip]{massHSB.eps}
\caption{Same as Fig.~\ref{fig:masshs}, but including the $\etmiss>$ 100 GeV cut.}
\label{fig:masshsB}}
\FIGURE[t]{
\includegraphics[width=11cm,clip]{massDR3B.eps}
\caption{Same as Fig.~\ref{fig:massdr3}, but including the $\etmiss>$ 100 GeV cut.}
\label{fig:massdr3B}}
\FIGURE[t]{
\includegraphics[width=11cm,clip]{meffB.eps}
\caption{Effective mass scale of events with $n(b)\ge 3$ after C1 cuts for the BG, points HSb 
and DRb, and two points on the HS and DR3 model lines with heavier gluinos 
($m_{\tilde g}=576$~GeV and $m_{\tilde g}=581$~GeV, respectively).}
\label{fig:meffB}}

\subsection{Multi-lepton channels}

In Fig.~\ref{fig:nmuonsB}, we show the isolated lepton multiplicity distribution
after cuts C1. In the $n(\ell )=0$ channel, the DR3b signal exceeds BG, while
in the $n(\ell )= 2$ channel, the HSb signal exceeds BG. In the $n(\ell) =1$ channel, 
BG from $t\bar{t}$ production is larger than both signal cases.
One can pull out a better signal rate in the $1\ell$ channel by imposing in addition
a $M_T(\ell,\eslt )>100$ GeV cut, as is well known\cite{bbkt}.

We divide the $n(\ell )=2$ channel into opposite-sign/same-flavor (OS/SF) events
($e^+e^-$ or $\mu^+\mu^- $), and same-sign events.
For the OS/SF channel, we see the $\etmiss$ cut has little effect on the signal-to-BG
ratio, since the background, mainly $t\bar{t}$ production, also has large
$\eslt$. 
Again, we expect the entire distribution to be self normalizing, 
since $t\bar{t}$ is the dominant BG, and one can fix the total $t\bar{t}$ 
cross section by normalizing to the $1\ell$ channel. Then the $n(\ell )$ distribution
should be harder than expected from just SM physics if a SUSY signal is present.
\FIGURE[t]{
\includegraphics[width=11cm,clip]{nmuonsB.eps}
\caption{ Lepton multiplicity distribution after C1
cuts at the LHC, with  $\sqrt{s}=7$ TeV. We show the signal
levels for the HSb (red) and DR3b (green) points along with
various SM backgrounds. In the $n(\ell )=2$
bin, the left (pluses) and right (crosses) columns show the background (black) and
signal components for OS/SF and SS dileptons.  }
\label{fig:nmuonsB}}

In Fig.~\ref{fig:invmassB}, we show the OS/SF dilepton invariant mass distribution
after cuts C1. 
Now, despite the negligible background, the reduction in the
signal makes the mass edges less visible. In particular, the dilepton
invariant mass still shows the $m_{\tz_2}-m_{\tz_1}$ edge, but with a smaller statistical
significance as evident from the error bars.
Performing a different-flavor subtraction may reduce the already negligible BG even further.
\FIGURE[t]{
\includegraphics[width=11cm,clip]{invmassB.eps}
\caption{OS dilepton ($\mu$'s and $e$'s) invariant mass for the HSb (red) and DR3b (green) points plus background events
after the C1 cuts (see text).
The BG distribution (gray) and the statistical error bars for 1 fb$^{-1}$ of integrated luminosity are also shown.
  }
\label{fig:invmassB}}

\subsection{Jets plus $Z\to\ell\bar{\ell}+\eslt$ signal}

We also show in Fig.~\ref{fig:invmassB} two additional HS and DR3 cases with $m_{\tg}\sim 525$ GeV. 
In this case, according to Fig.~\ref{fig:bfz2}, the value of $m_{\tz_2}$ is high enough that the two-body decay
$\tz_2\to\tz_1 Z$ is now dominant. We then expect a signature of multiple $b$-jets plus $\eslt$ plus
a dilepton pair which reconstructs to $m(\ell^+\ell^- )\simeq M_Z$\cite{btw}. While signal rates can be very high 
for this channel, SM background is low, coming from processes such as $t\bar{t}$ and $Z+t\bar{t}$.

\subsection{LHC reach for Yukawa-unified SUSY using $\eslt$ and $e$ ID}
\label{sec:reach}

Next, we investigate the full LHC reach for Yukawa-unified SUSY along the HS and DR3 model lines.
First, we require the cut set C1, which includes $\eslt>100$ GeV, and then require
$n(b)\ge 2$ or $n(b)\ge 3$. The SM background level and $5\sigma$ level for
0.2 and 1 fb$^{-1}$ are shown in the plots of Fig.~\ref{fig:reach_b}, 
along with expected signal rates from the HS and DR3 model lines.
For the $n(b)\ge 2$ case, we find an LHC reach for Yukawa-unifed SUSY out to
$m_{\tg}=500$ (600) GeV for 0.2 (1) fb$^{-1}$. 
The reach is largely independent of whether one is in the HS or the DR3 model.

For $n(b)\ge 3$, the SM background
is greatly reduced. In this case, we find a reach to $m_{\tg}=540$  (630) GeV 
for 0.2 (1) fb$^{-1}$. Again, the reach in the multi-$b$-jet channel is largely independent
of the model line, since both give large numbers of $b$-jets + $\etmiss$ in the final state.
We have also calculated the reach in the $n(b)\ge 4$ channel; here, the result is 
qualitatively similar to that obtained in the $n(b)\ge 2$ case.
\FIGURE[t]{
\includegraphics[width=11cm,clip]{mg22bB.eps}
\includegraphics[width=11cm,clip]{mg23bB.eps}
\caption{Early LHC reach for Yukawa-unified SUSY using cuts $C1$ plus 
$n_b\ge 2$ and $n_b\ge 3$.
}
\label{fig:reach_b}}

Next, in Fig.~\ref{fig:reach_l}, we show the LHC reach for Yukawa-unified SUSY using the $C1$ cuts plus
requiring a pair of OS/SF dileptons. In this case, the LHC reach is model dependent. For the
HS model, where we obtain a high rate for $\tg\to b\bar{b}\tz_2$ decays, we find a reach up 
to $m_{\tg} =400$ (500) GeV for 0.2 (1) fb$^{-1}$ of integrated luminosity. For the DR3 model line,
there is no reach for 0.2 fb$^{-1}$, since here the $\tg\to b\bar{b}\tz_1$ decay is dominant. 
With 1~fb$^{-1}$, however, a signal with $5\sigma$ significance should be visible for 
$m_{\tg}\sim 300-450$ GeV.
\FIGURE[t]{
\includegraphics[width=11cm,clip]{mg2OSB.eps}
\caption{Early LHC reach for Yukawa-unified SUSY using cuts $C1$ 
for events containing OS/SF dileptons.
}
\label{fig:reach_l}}

For $m_{\tg}\agt 450$ GeV, the two body decay mode $\tz_2\to\tz_1 Z$ opens up and
we expect to reconstruct $Z\to e^+e^-$ and $Z\to \mu^+\mu^-$ within the class of signal
events. Here, we will adopt cuts C1, but in addition require a OS/SF dilepton pair
with 75 GeV$<m(\ell^+\ell^- )<$ 105 GeV (cuts $C1'$). For this topology--
$\ge 4$ jets $+\eslt +Z\to\ell^+\ell^-$-- the dominant SM BG comes
from $t\bar{t}$ production. 
The LHC reach is shown in Fig.~\ref{fig:reach_z}. 
As can be seen, no significant excess is expected with $0.2~\rm fb^{-1}$ of data. 
However, for $\agt 1$ fb$^{-1}$, we find that this topology can produce a $5\sigma $ signal 
for $m_{\tg}\sim450-530$ GeV for both the HS and DR3 model lines.
\FIGURE[t]{
\includegraphics[width=11cm,clip]{mg2OSC.eps}
\caption{LHC reach for Yukawa-unified SUSY using cuts $C1$' 
with a reconstructed leptonic $Z$ boson.
}
\label{fig:reach_z}}

\subsection{Differentiating the HS and DR3 models}
\label{sec:diffmodel}

Here we discuss the possibility of distinguishing the two models discussed so far
using low luminosities and the channels investigated in the previous sections. As seen from
last section results (Figs.~\ref{fig:reach_b} and \ref{fig:reach_l}) and the discussion in Sec. \ref{sec:pheno},
for low to moderate $m_{\tg}$ masses we expect the HS and DR3 models to have rather distinct signatures.
The first one is rich in multi-b jets ($n(b) \ge$ 4, 5) and OS/SF dileptons coming from the $\tg \to \tz_2 + b\bar{b}$
followed by $\tz_2 \to Z/Z^* + b\bar{b}/\ell^+\ell^-$ decays, but has a softer $\eslt$ spectrum due to the two step
cascade decays. On the other hand, the DR3 model is mainly dominated by $\tg \to \tz_1 + b\bar{b}$, with
small cross-sections in the multi-lepton channels, a moderate number of $b$-jets ($n(b) \ge$ 3,4) and 
a harder $\eslt$ spectrum. To explore these features and discuss how well we can distinguish both models with year one
data, we plot in Fig.~\ref{fig:diff} the ratio of the OS and $n(b)\ge 3$ channels. As expected from the above discussion,
we see that the ratio is larger for the HS model and suppressed in the DR3 case for $m_{\tg} \lesssim 600$ GeV.
As the gluino mass increases, the $\tw_1 + bt$ and $\tz_1 + b\bar{b}$ gluino decay 
channels become available (see Fig.~\ref{fig:brs}),
increasing the OS channel in both models. From Fig.~\ref{fig:diff} we see that the OS/3b ratio should be a good discriminator
up to $m_{\tg} \sim 450$ GeV. For $m_{\tg} \gtrsim 450$ GeV, higher luminosities are required in order to distinguish the models.
But with high enough luminosities, we should be able to tell the models apart up to $m_{\tg} \sim 600$ GeV. For even higher
$m_{\tg}$ values, the signal features are too similar and the simple OS/3b ratio is no longer useful.
\FIGURE[t]{
\includegraphics[width=11cm,clip]{OSvs3b.eps}
\caption{The ratio of the OS and $n(b)\ge 3$ cross-sections after C1 cuts for the HS and DR3 model
lines plus background as a function of the gluino mass. We also show the statistical error bars
for 1 fb$^{-1}$.
}
\label{fig:diff}}

\clearpage

\section{Comparison with CDF/CMS multijets + $\eslt$ channel}
\label{sec:cmscdfjetmet}

Next we would like to compare our results for the multi-$b$-jets + $\eslt$ signature with a multijets + $\eslt$ analysis for SUSY searches proposed and used by CDF~\cite{cdfjetmet} and further developed in CMS~\cite{SS_thesis}.   We implement the following selection.
\bi
\item $n(jets)\ge 3$ (we take jet $p_T>50$~GeV),
\item $\eslt\ge 150$ GeV,
\item $\Delta\phi(\vec{E}^{miss}_T,\vec{H}_T)<1$,
\item $0.5< R_2\equiv\sqrt{\Delta\phi^2(j1,\eslt )+(\pi-\Delta\phi(j2,\eslt))^2}<4$,
\item $1.5< R_{a1}\equiv\sqrt{(\Delta\phi(j2,\eslt ))^2+(\Delta\phi(j3,\eslt))^2}<4.25$,
\item $\Delta\phi (j2,\eslt )>0.35$,
\item $\Delta\phi (j_i ,\eslt )>0.3$,
\item $|\eta (j1)|<1.7$ ,
\item Finally, we require the highest $E_T$ object in each signal event to be hadronic, rather than leptonic~\footnote{This is an approximation for the leading track isolation step of the indirect lepton veto (ILV) technique included in the CDF/CMS selection.  Normally in multijets $\eslt$ analyses where lepton information is not explicitly used, $W/Z/t\bar{t}+n$~jets backgrounds with leptonic $W,Z$ decays can be eliminated by vetoing the events whose leading track is isolated.  We omit the second step of ILV featuring jet electromagnetic and charged fractions since it is designed to eliminate machine and cosmic backgrounds which we do not consider here.}.
\ei
The angular cuts are introduced in multijets + $\eslt$ analyses where no lepton information is explicitly used in order to guarantee the discrimination of QCD backgrounds which may have large $\eslt$ arising primarily due to jet mismeasurements.  
In such events, $\eslt$ typically aligns with the 2nd hardest jet, as the hardest jet has a tendency to be mismeasured most and eventually becomes the 2nd hardest jet.  

The reach results for CDF/CMS multijets + $\eslt$ cuts with no $n(b)$ requirement 
can be seen in Fig.~\ref{fig:reach_cdfcmsjetmet}.
The $5\sigma$ reach for 0.2 (1) fb$^{-1}$ of integrated luminosity is found to be $m_{\tg}\sim 370$ (430) GeV. 
\FIGURE[t]{
\includegraphics[width=11cm,clip]{mg2Sezen.eps}
\caption{LHC reach for Yukawa-unified SUSY using CDF/CMS multijets + $\eslt$ cuts with
no $n(b)$ requirement.
}
\label{fig:reach_cdfcmsjetmet}}

In Figure~\ref{fig:nbjetscdfcms}, we plot the $n(b)$ signal and background distributions after CDF/CMS cuts, 
which share similar characteristics with the distributions after C1 cuts.  
The dominance of signal starting with $n(b)=1$ illustrates the importance of considering $b-$jet tagging in 
the multijets + $\eslt$ analyses.  The original CMS analysis made use of the $B-$triggers to significantly 
enhance the signal/BG ratio, while here we directly cut on $n(b)$.  We show in Table~\ref{table:cdfcmsjetmet} 
the cross sections found after implementing CDF/CMS cuts with no requirement on $n(b)$ and with $n(b)=1,2,3$.
\FIGURE[t]{
\includegraphics[width=11cm,clip]{nbjets_Sezen.eps}
\caption{ $b$-jet distribution after CDF/CMS multijets + $\eslt$ 
cuts at the LHC, with  $\sqrt{s}=7$ TeV. We show the signal
levels for the HSb (red) and DR3b (green) points along with
various SM backgrounds.
  }
\label{fig:nbjetscdfcms}}

\begin{table}
\centering
\begin{tabular}{|c|c|c|c|c|}
\hline
\multicolumn{5}{|c|}{Results after CDF/CMS multijets + $\eslt$-based selection} \\
\hline
 & $\sigma$(no $n(b)$ req.) & $\sigma$($n(b)\geq$ 1)  &$\sigma$($n(b)\geq$ 2) &$\sigma$($n(b)\geq$ 3)  \\
\hline
HSb  & 390 fb & 313 fb  & 167 fb & 55 fb \\ \hline
DR3b  & 849 fb & 739 fb & 435 fb & 140 fb\\ \hline
BG & 1132 fb & 366 fb & 101 fb & 7 fb\\
    \hline
\end{tabular}
\caption[]{Cross-sections for the channels with no $n(b)$ requirement, $n(b)\geq $ 1, 2 and 3 
after the CDF/CMS multijets + $\eslt$ cuts for the points HSb, DR3b and the background. }
\label{table:cdfcmsjetmet}
\end{table}

As before, we can do much better by requiring a high multiplicity of $b$-jets. 
In Fig.~\ref{fig:reach_cdfcmsjetmetB}, we adopt the same cuts as in Fig.~\ref{fig:reach_cdfcmsjetmet}, 
but in addition require $n(b)\ge 3$. 
In this case, the BG drops by a factor of $\sim 150$, while the signal drops merely by a factor of 
$\sim 7$ for $m_{\tg}\sim 300$ GeV. 
The LHC reach increases to $m_{\tg}\sim 400$ (500) GeV for 0.2 (1) fb$^{-1}$ of integrated luminosity.
\FIGURE[t]{
\includegraphics[width=11cm,clip]{mg2SezenB3.eps}
\caption{LHC reach for Yukawa-unified SUSY using CDF/CMS multijets + $\eslt$ cuts with
$n_b\ge 3$.
}
\label{fig:reach_cdfcmsjetmetB}}

\section{Conclusions}
\label{sec:conclude}

In $t-b-\tau$ Yukawa-unified SUSY, we expect a characteristic spectrum of superpartners
with first/second generation squarks and sleptons around 10 TeV, third generation sparticles, 
heavy Higgs bosons and $\mu$ around the few TeV level, and very light gauginos, with 
$m_{\tg}\sim 300-500$ GeV (although here we consider even higher values).
Thus, at LHC, we expect to see gluino pair production at a high rate, followed by gluino
decays to $b\bar{b}\tz_i$ or $t\bar{b}\tw_1^- +c.c.$. 
SUSY searches should therefore exploit the high multiplicity of $b$-jets expected in this 
scenario.

We investigated two model lines-- the HS and DR3 cases. The HS case leads to large rates for
OS/SF dileptons in the final state, while DR3 case does not.
We computed numerous $2\to 2$, $2\to 3$ and $2\to 4$ background processes. We found that
with just 0.1--0.2~fb$^{-1}$ of integrated luminosity, the LHC discovery reach with $\sqrt{s}=7$ TeV
extends out to $m_{\tg}\sim 400$ GeV, even without using $\eslt$ cuts. Crucial use is made of 
the high multiplicity of $b$-jets in the final state. In the case of the HS model, 
a corroborating signal appears in the $\mu^+\mu^- +jets +\ge 1$ $b$-jet channel.

The LHC reach at very low luminosity and without $\eslt$ is comparable to 
the Tevatron reach in the multi-$b+\eslt$ channel with 5--10~fb$^{-1}$ 
of data. This may lead to a tight competition for the discovery or exclusion 
of the simplest Yukawa-unified SUSY scenario!

Moving beyond about 0.2~fb$^{-1}$, we expect reliable $\eslt$ resolution and electron identification 
to become available.
We find that the LHC reach, using $\sqrt{s}=7$ TeV and 1 fb$^{-1}$ of integrated luminosity, 
will move into the $m_{\tg}\sim 600-650$ GeV
range for both the HS and DR3 model lines, if we require $\eslt \ge 100$ GeV along with 
$n(b)\ge 3$. This reach is presumably sufficient to rule out
Yukawa unification in the DR3 case. 
In the HS case, somewhat larger values of $m_{\tg}$ can be allowed, although they seem
very improbable.
The LHC reach for 1 fb$^{-1}$ using the multi-jets + $\eslt$ signature-- but making no
requirement on $n(b)$-- turns out to be much lower.
A summary of our various results is presented in a convenient form in Table \ref{tab:reach}.
\begin{table}
\centering
\begin{tabular}{|c|c|c|c|c|c|}
\hline
& $\mathcal{L}$ (fb$^{-1}$) & 0.05 & 0.1 & 0.2 & 1 \\
\hline
\multirow{4}{*}{C0} & HS & 340 GeV & 371 GeV & 400 GeV & 471 GeV\\ 
& \small \textit{Channel} & \small $n(b) \geq 3$ & \small $n(b) \geq 3$ & \small $n(b) \geq 3$ & \small $n(b) \geq 4$\\ 
& DR3 & 340 GeV & 363 GeV & 394 GeV & 469 GeV \\ 
& \small \textit{Channel} & \small $n(b) \geq 3$ & \small $n(b) \geq 3$ & \small $n(b) \geq 3$ & \small $n(b) \geq 3$ \\ \hline
\multirow{4}{*}{C1} & HS & 436 GeV & 480 GeV & 526 GeV & 630 GeV \\  
&  \small \textit{Channel} & \small $n(b) \geq 3$ & \small $n(b) \geq 3$ & \small $n(b) \geq 3$ & \small $n(b) \geq 3$ \\ 
& DR3 & 460 GeV & 506 GeV & 545 GeV & 630 GeV \\ 
&  \small \textit{Channel} & \small $n(b) \geq 3$ & \small $n(b) \geq 3$ & \small $n(b) \geq 3$ & \small $n(b) \geq 3$ \\ \hline
\multirow{4}{*}{CDF/CMS} & HS & - & 341 GeV & 380 GeV & 474 GeV \\  
&  \small \textit{Channel} & \small - & \small $n(b) \geq 1$ & \small $n(b) \geq 1$ & \small $n(b) \geq 3$ \\ 
& DR3 & 350 GeV & 382 GeV & 420 GeV & 516 GeV \\ 
&  \small \textit{Channel} & \small $n(b) \geq 2$ & \small $n(b) \geq 2$ & \small $n(b) \geq 2$ & \small $n(b) \geq 3$ \\ \hline
\end{tabular}
\caption{ Gluino $5\sigma$ mass reach for the HS and DR3 model lines for different luminosites. 
We show the values for the early search (C0 cuts), the full reach (C1 cuts) and the CDF/CMS 
multijets + $\eslt$ cuts (CDF/CMS). We also show the channel which optimizes the reach.}
\label{tab:reach}
\end{table}

At some point LHC energy will be increased to $\sim 10$ TeV, and this will allow the reach to be
extended past the values presented here. 
Thus, our main conclusion is that LHC stands an excellent chance to either discover Yukawa-unifed
SUSY during year 1 of operation, or exclude almost all its model parameter space!

Last but not least we note that, if the scenario discussed here is realized, 
a trilepton signal from Drell-Yang $\tw_1\tz_2$ production should appear when moving into the 
several fb$^{-1}$ regime, giving direct access to the chargino/neutralino sector. 

{\it Note added:} While finalizing this manuscript, we found an Atlas note\cite{atlas} 
which also investigates
using multiple $b$-jets to enhance the LHC reach for SUSY in the mSUGRA model. They 
examine integrated luminosity values of 0.1 and 1 fb$^{-1}$, but take $\sqrt{s}=14$ TeV.
Their overall results seem quite consistent with the results given here.

\acknowledgments
We thank Harrison Prosper for sharing with us his p-value and significance code.  
This research was supported in part by the U.S. Department of Energy grant number 
DE-FG-97ER41022, by the Fulbright Program and CAPES (Brazilian Federal Agency 
for Post-Graduate Education).
The work of SK is supported in part by the French ANR project ToolsDMColl, 
BLAN07-2-194882.
	
%

\end{document}